\documentclass[prb,twocolumn,showpacs,superscriptaddress]{revtex4}
\usepackage{amsfonts}
\usepackage{epsfig}
\usepackage{bm}
\usepackage{amsmath}

\begin{document}

\title{Numerical study of the localization length critical index
in a network model of plateau-plateau transitions in the quantum Hall effect}

\author{M. Amado}

\affiliation{GISC, Departamento de F\'{\i}sica de Materiales, Universidad
Complutense, E-28040 Madrid, Spain}

\author{A. V. Malyshev}

\affiliation{GISC, Departamento de F\'{\i}sica de Materiales, Universidad
Complutense, E-28040 Madrid, Spain}

\author{A. Sedrakyan}

\affiliation{GISC, Departamento de F\'{\i}sica de Materiales, Universidad
Complutense, E-28040 Madrid, Spain}

\affiliation{Yerevan Physics Institute, Br. Alikhanian 2, Yerevan 36, Armenia}

\author{F. Dom\'{\i}nguez-Adame}

\affiliation{GISC, Departamento de F\'{\i}sica de Materiales, Universidad
Complutense, E-28040 Madrid, Spain}

\begin{abstract} 

We calculate numerically the localization length critical index within the
Chalker-Coddington (CC) model for plateau-plateau transitions in the quantum
Hall effect. Lyapunov exponents have been calculated with relative errors on
the order $10^{-3}$. Such high precision was obtained by considering the
distribution of Lyapunov exponents for large ensembles of relatively short
chains and calculating the ensemble average values. We analyze thoroughly
finite size effects and find the localization length critical index $\nu=
2.517\pm 0.018$.

\end{abstract}


\maketitle

Plateau-plateau transitions in the quantum Hall effect have been one of the
most challenging problems in condensed matter physics during the last two
decades. It is a interesting example of the localization-delocalization
transition in two-dimensional disordered systems where a quantum critical
point appears due to the breaking of time reversal symmetry. The most
important problem in this area of research is the formulation of a quantum
field theory describing the transition. The first suggestion in this respect
appeared in Ref.~\onlinecite{LLP}, where the authors noticed that the
presence of the topological term in the nonlinear sigma model formulation of
the problem can result in the occurrence of delocalized states in strong
magnetic fields.

Later, Chalker and Coddington~\cite{CC} formulated a phenomenological model
of quantum percolation based on a transfer matrix approach (refered to as
the CC model hereafter). The numerical value $2.5\pm 0.5$ of the critical
index of the Lyapunov exponent calculated within the CC model (see
Ref.~\onlinecite{BH} for a review) was in good agreement with the
experimentally measured correlation length index $\nu= 2.4$ in the quantum
Hall effect.\cite{Wei94} This success motivated considerable interest in the
CC model and stimulated its further investigation until present
days.~\cite{LWK,DHL1,DHL2,ChoF1,Zirn1,AS,Tsvelik,LC,SO} In early
studies~\cite{LWK,DHL1,DHL2} the continuum limit of the CC model was related
to supersymmetric spin chains, which was further developed by
Zirnbauer.~\cite{Zirn1} In Refs.~\onlinecite{Tsvelik} and~\onlinecite{LC}
the continuum limit was also related to the conformal field theory of
Wess-Zumino-Witten-Novikov (WZWN) type. Analyzing the representations of the
$\mathrm{PSL}(2|2)$ conformal field theory, they found one which gives a
reasonable value of $16/7\simeq 2.286$ for the correlation length index.
Moreover, multi-critical scaling indices of the CC model were predicted to
depend quadratically on the level $q$ of multi-criticality within the WZWN
model approach. 

Most intriguing developments in the plateau-plateau
transition problem were reported later,~\cite{Gruzberg,Evers} where the
multi-critical behavior of the CC model was investigated. In both papers
quartic dependence of the multi-critical indices of the parameter $q$ was
observed, in contrast to earlier predictions.~\cite{Tsvelik,LC} The latter
suggested that the validity of the supersymmetric WZWN approach to
plateau-plateau transitions in the quantum Hall effect is questionable. On
the other hand, since the plateau-plateau transitions are of the second
order, there is a conformal symmetry at the transition point and there
should exist a conformal field theory describing it. Candidate theories
could be tested against the experimental data by comparing critical indices.
Unfortunately, the precision of the available experimental indices is too
low and does not enable us to identify the correct theory. Therefore,
comparison to numerically calculated values can be more feasible for this
task. However, reliable calculation of the localization length critical
index with good precision has been known to be a very challenging task and
there is still little consensus on the obtained values and especially their
error bars. This paper is largely motivated by the demand of such an
accurate calculation.

We have carried out numerical calculations of the Lyapunov exponent and the
corresponding critical index in the CC model taking finite size effects into
account. Our calculation strategy can be summarized as follows. Instead of
finding the eigenvalues of the product $T=\prod_{j=1}^L T_j$ of a large
number of transfer matrices $T_j$, we form a large ensemble of $L_r$ shorter
products and obtain the distribution function of the smallest Lyapunov
exponent $\gamma= (1/2L)\ln \lambda$, where $\lambda$ is the eigenvalue of
the matrix $S=T^\dagger T$ closest to the unity from above. Then we
calculate the mean value and the standard deviation of the distribution.
According to the central limit theorem the distribution of the mean value is
normal~\cite{Tut}. This procedure allows us to effectively increase the
product length up to $L_\textrm{eff}=L_r\times L$ and reduce the error,
which is on the order of $\sqrt{\gamma/L_\textrm{eff}}$. The error can
therefore be reduced by collecting the statistics of eigenvalues. In finding
the correct value of the critical exponent we take into account the finite
size of matrices $T$ and do the standard finite size scaling (FSS) analysis
according to Refs.~\onlinecite{Kramer, Ohtsuki, Roemer}. The result we found
for the correlation length index is $\nu= 2.517\pm 0.018$.

\begin{figure}[t]
\centerline{\includegraphics[width=75mm,angle=0,clip]{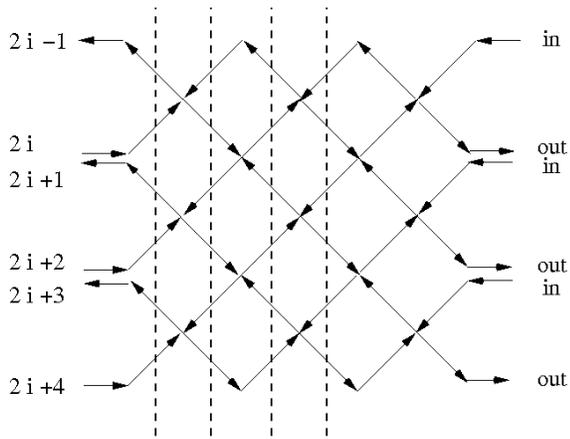}}
\caption{Schematic view of the CC model network structure.}
\label{fig:F0}
\end{figure}

The transfer matrix of the CC model consists of local $2\times 2$ matrices
of the form
\begin{eqnarray}
\left( \begin{array}{c}
w_{1 R} \\
w_{2 R}
\end{array}
\right)=
\left(
\begin{array}{cc}
1/t & r/t \\
r/t & 1/t \\
\end{array}
\right)
\left( \begin{array}{cc}
e^{i \alpha_1} & 0 \\
0 & e^{i \alpha_2}\\
\end{array}
\right)
\left( \begin{array}{c}
w_{1 L} \\
w_{2 L}
\end{array}
\right)\ ,
\label{CCTr}
\end{eqnarray}
where $t=1/\sqrt{1+e^{2x}}$ is the tunneling amplitude of electronic states
and $r=\sqrt{1-t^2}$ is the rotation at each node of the regular lattice
shown in Fig.~\ref{fig:F0}. The model parameter $x$ corresponds to the Fermi
energy while phases $\alpha_{i}$ ($i=1,2$) are random variables from $0$ to
$2\pi$, arising from the randomness of the potential landscape. $N$ of such
local transfer matrices correspond to a vertical strip in Fig.~\ref{fig:F0}
that forms $2 N \times 2 N$ transfer matrices $W_1 U_1$ and $W_2 U_2$ for
neighboring columns. The parameterization of transfer matrices $W_1$ and
$W_2$ in neighboring columns is such that it is invariant under the rotation
by $\pi/2$, the symmetry which is obvious from Fig.~\ref{fig:F0}. We
therefore have two parameters $x$ and $x^{\prime}$ corresponding to
neighboring columns related as follows
\begin{equation}
\label{02}
t=r^{\prime}=\frac{1}{\sqrt{1+e^{2x}}},\;\;r=t^{\prime}=
\frac{1}{\sqrt{1+e^{2x^{\prime}}}}\ .
\end{equation}
The transition point $x_c$ is equal to $0$.  As the result, matrix elements
of $W_1$ and $W_2$ are defined as follows:
\begin{align*}
\label{WW}
\left[W_1\right]_{2n+1\,2n+1}=\left[W_1\right]_{2n\,2n}&=1/t\ ,\nonumber\\
\left[W_1\right]_{2n+1\,2n}= \left[W_1\right]_{2n\,2n+1}&=r/t\ ,\nonumber\\
\left[W_2\right]_{2n-1\,2n-1}=\left[W_2\right]_{2n\,2n}&=1/t^{\prime}\ ,\\
\left[W_2\right]_{2n-1\,2n}=\left[W_2\right]_{2n\,2n-1}&=r^{\prime}/t^{\prime}\ ,\nonumber\\
\left[W_2\right]_{1\,N}=\left[W_2\right]_{N\,1}&=r^{\prime}/t^{\prime}\ ,\nonumber
\end{align*}
where $n=1 \dots N$ and periodic boundary conditions are imposed on $W_2$.
Matrices $U$ have a simple diagonal form
$\left[U_{1,2}\right]_{nm}=\exp{(i\alpha_n)}\,\delta_{nm}$. The final
transfer matrix $T$ can then be represented as a product $\prod_{j=1}^L W_1
U_{1j}W_2 U_{2j}$.

The largest matrix element of the transfer matrix $T$ is of the order of
$\eta\sim(1+e^{2x})^{L}$, growing exponentially with the system length
$L$.\footnote{the estimate is made with the exponential precision, the
largest matrix element grows faster because of a polynomial pre-factor which
is difficult to derive.} Because we are looking for eigenvalues of
$S=TT^\dagger$ which are closest to unity, we need to compute $S$ with the
precision of about $2\log_{10}\eta$, otherwise numbers of the order of unity
are unreliable due to round-off errors. We used arbitrary precision
calculations with adaptive precison increasing it gradually for each matrix
product. We calculated the transfer matrix $T$ of the system of length $L$
on a grid $\{N,x\}$ using a graph-like algorithm. Instead of computing
product of local transfer matrices sequentially, we calculate it on a full
$k$-way tree with adaptive precision. At each node we estimated the required
precision and set the involved matrices to it before multiplying them. In
all calculation we used $k=7$, which allowed us to perform the vast majority
of the matrix multiplications (at the leaf node level) with the machine
precision. When the matrix $S=TT^\dagger$ was computed we used the
shift-and-invert Lanczos algorithm to calculate a set of selected
eigenvalues of $S$ which are closest to unity. Typically, we calculated six
eigenvalues with the accuracy of $10^{-5}$, checking that these eigenvalues
appeared in pairs and satisfied $\log{\lambda_{2i}}=-\log{\lambda_{2i+1}}$
(where $i=0,1,2$  and $\lambda_1$ is the eigenvalue closest to unity from
above, i. e. $\lambda_1>1$). We then calculated Lyapunov exponents
$-(1/2N)\log[\lambda_i(N,x)]$. Hereafter, we only focus on the smallest
Lyapunov exponent $\gamma_N(x)$.

\begin{figure}[t]
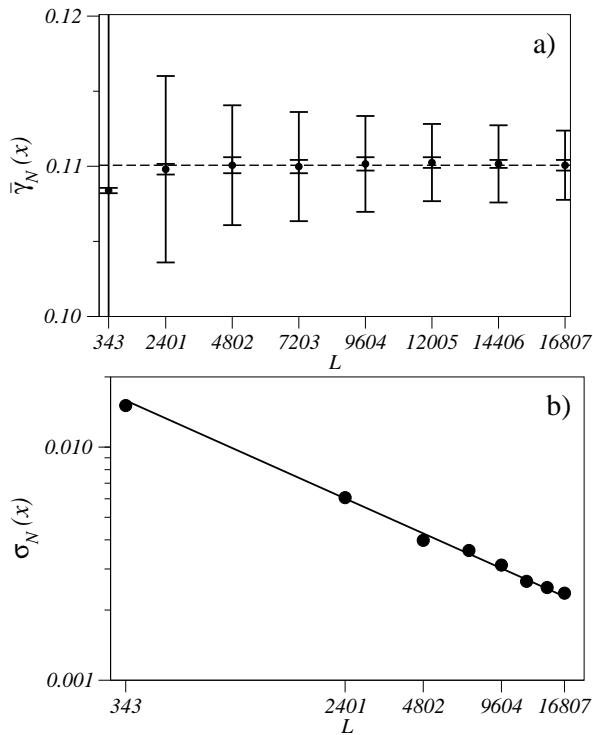

\begin{center}
	\includegraphics[width=.9\columnwidth,clip]{figure2a.eps}\\
	\includegraphics[width=.9\columnwidth,clip]{figure2b.eps}
\end{center}
\caption{
a)~Mean values of the $\gamma_N(x)$-distribution and estimated error bars
calculated for $N=60$, $x=0.404$ at different node levels ($L=343, 2401,
\ldots, 16807$). The horizontal dashed line is a guide for the eye, showing
the mean value for $L=16807$. Larger error bars show the standard deviation
of the distributions $\sigma_N(x)$, while smaller error bars are equal to
$\bar\sigma_N(x)$ and estimate the error of the mean value. b)~Full circles
show the standard deviation $\sigma_N(x)$ as a function of the system length
$L$ and the solid line is the best $1/\sqrt{L}$ fit to the data.
}
\label{gammaMdist}
\end{figure}

The tree-like algorithm provides access to partial transfer matrices and
allows us to calculate Lyapunov exponents at different node levels, not only
at the root node. The latter offers the possibility to collect and study the
statistics of Lyapunov exponents. We analyzed distributions of $\gamma_N(x)$
for different system lengths $L$. Figure~\ref{gammaMdist}(a) displays the
mean value $\bar\gamma_N(x)$ of the distribution and its estimated standard
errors calculated for $N=60$, $x=0.404$, and $L=343, 2401, \ldots, 16807$.
Larger error bars show the standard deviation $\sigma_N(x)$ of the
distribution at each node (i.e., at each $L$). Smaller error bars represent
the standard error of the mean value itself, estimated as
\begin{equation}
\bar\sigma_N(x)=\sigma_N(x)/\sqrt{L_r}\ ,
\label{error}
\end{equation}
where $L_r$ is the size of the Lyapunov exponent ensemble at each node.
Figure~\ref{gammaMdist}(b) demonstrates that the standard deviation
$\sigma_N(x)$ varies as $1/\sqrt{L}$ with the system length, confirming the
applicability of the central limit theorem and, hence, the
definition~(\ref{error}).

Figure~\ref{gammaMdist} shows that values of the disorder-averaged mean
Lyapunov exponent for $N \geq 2401$ are the same within their error bars. We
calculated these quantities for different sets of $N$ and $x$, as well as
for longer chains, and found the same result. Therefore, rather than
calculating the exponent for a very long chain, we can calculate the mean
Lyapunov exponent for a large number $L_r$ of relatively short chains, which
is more efficient. Varying $L_r$ we can achieve any given target accuracy;
in our calculation we used $L=4802$ with the typical ensemble size of
$N_r\approx2000$.

Having calculated disorder-averaged Lyapunov exponents, we use the standard
FSS analysis of the data, formulated in Ref.~\onlinecite{Kramer} and
extended in the Refs.~\onlinecite{Ohtsuki,Roemer,Eilmes}, in order to obtain
the localization length index $\nu$. The Lyapunov exponent is believed to
have scaling behavior (see for example Refs.~\onlinecite{Ohtsuki}
and~\onlinecite{SO} and references therein), and finite size effects can be
accounted for by the following formula for the scaling function
$\Gamma(N,x)$ which approximates $N\bar\gamma_N(x)$ in the vicinity of the
critical point
\begin{eqnarray}
\label{fss}
\Gamma(N,x)=F_0(N^{1/\nu} u_0)+F_1(N^{1/\nu} u_0)N^y\, u_1
\end{eqnarray}
%
where $F_0(\cdot)$, $F_1(\cdot)$, $u_0(\cdot)$ and $u_1(\cdot)$ are
universal, independent of $L$ functions. The first function $F_0(\cdot)$ is
the contribution of the main operator in the corresponding conformal field
theory, which defines the correlation length. The second function
$F_1(\cdot)$ results from the operator with the anomalous dimension which is
close to that of the main one. We used the following formula to fit to the
data\cite{Eilmes,SO}:
\begin{equation}
 \label{func2fit}
\Gamma(N,x) = \sum_{n=0}^{3} a_{0\,2n}\left[u_0(x)\,N^{1/\nu}\right]^{2n}
+ a_{10}\, u_1(x)\,N^{y}
\end{equation}
where $u_0(x)=x+b_3x^3$ and $u_1(x)=1+c_2x^2$, so the series for $F_0$,
$F_1$, $u_0$, and $u_1$ were truncated at orders 6, 0, 3 and 2,
respectively. Further increase of these orders turned up to be impractical
as it deteriorated the result.

\begin{figure}[t]
\begin{center}
\includegraphics[width=0.97\columnwidth,clip]{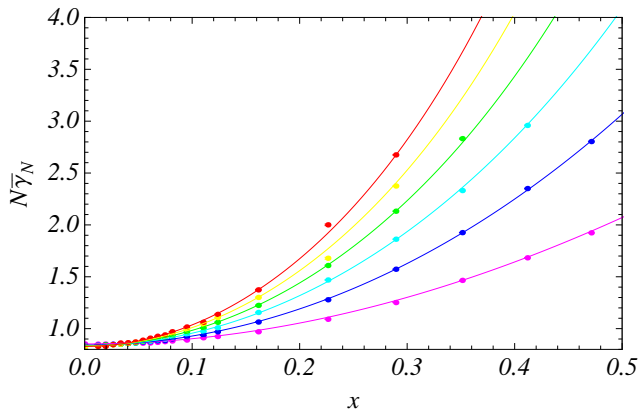}
\end{center}
\caption{
Simpulated data and the fit of Eq.~\ref{func2fit} to them ($c_1=0$). From
bottom to top (for $x>0.2$): $N=10, 20, \dots, 60$.}
\label{best-result}
\end{figure}

Because all mean Lyapunov exponents have different error bars we used a
weighted fit with $[N\bar\sigma_L(x)]^{-2}$ as weights. We also put the
following constraints on the parameters: $0.5<a_{00}<1$, $|a_{0\,2n}|,
|a_{10}|, |b_{3}|,|c_{2}| \leq 2$, $2.1 \leq \nu \leq 2.7$ and $-0.5 \geq y
\geq 0$. We checked that the result is not affected by the constraints
while they stabilize the procedure and improve the convergence of the
nonlinear fit. All fits which converged to a constraint boundary were
discarded. Global optimization methods such as the simulated annealing were
used to find the best fit. The result of such a fit is presented in
Fig.~\ref{best-result}. All curves intersect at $x\approx0.025$; note that
the ordering of the curves at $x=0$ suggests an additional constraint,
namely $a_{10}>0$.

In order to estimate error bars of the parameters we used the standard
resampling technique.\cite{NR} Up to $1000$ generated synthetic data sets of
mean Lyapunov exponents were drawn from the corresponding normal
distributions centered at $\bar\gamma_N(x)$ and having the standard
deviation $\bar\sigma_N(x)$. Fitting to these data we obtained distributions
of the parameters and calculated their mean values and standard errors. The
following parameter values were found: for the case $c_1\neq0$ we obtained
$\nu=2.519\pm0.023$, $y=-0.32$ and the typical goodness of fit parameter was
$\chi^2=864.5$ for $107$ points. For $c_1=0$, $\nu=2.517\pm0.018$,
$y\approx-0.24$ and $\chi^2=848.2$. We note that the parameter $y$ is always
small, which suggests that finite size effects are very pronounced in the CC
model.

The curves for Lyapunov exponents for different sizes $N$ are intersecting
in a region close to $x=0$. The crossing point of a pair of such curves
corresponding to consecutive values of $N$ shifts towards the origin on
increasing $N$. The latter is the consequence of the condition $a_{10}>0$,
which increases the role of the next to the leading operator in the problem
for small $N$. This fact stresses the importance of the second operator
together with the main one in the analysis of possible candidate conformal
field theories which describe plateau-plateau transitions in the quantum
Hall effect.

The obtained value of the critical index $\nu=2.517\pm0.018$ disagrees with
early results,\cite{BH, Raikh, RC} which could be attributed to the quality
of the data, system sizes reached, and finite size effects. Those effects
prooved to be extremely important for accurate calculation of the critical
exponent even in the standard Anderson model,\cite{Ohtsuki} in which case
the irrelevant exponent $y\sim-3$ is much larger than in the CC model
(finite size effects are therefore less pronounced). On the other hand, our
critical exponent is smaller than the one calculated by Slevin and
Ohtsuki,\cite{SO} who obtained $\nu=2.593\pm0.006$. However, both high
precision numerical results are considerably above the experimental value of
$\nu=2.38 \pm 0.06$ measured recently in GaAs-AlGaAs
heterostructures.~\cite{Li1, Li2} The latter fact emphasizes the necessity
to investigate the validity of the CC model further.

In summary, the obtained critical exponent $\nu > 2.5$ suggests that the
rational value $7/3$ which is in agreement with some early calculations,
could be questioned. Our technique is based on the calculation of the
product of transfer matrices preserving the required precision during
iterations and collecting large statistics to achieve a given target
precision. Finite-size effects turn out to be very pronounced within the
framework of the CC model and should be taken into account properly in order
to obtain reliable results. In particular, the calculated value of the
critical exponent suggests that some WZWN-type models based on the conformal
field theory should be reconsidered, which demands new developments and
approaches in the formulation of the continuum limit of the CC model as well
as further studies and more accurate calculations of Lyapunov indices.

AS acknowledges discussions with I.\ Gruzberg, V.\ Kagalovsky and thanks for
hospitality the Universidad Complutense de Madrid where the major part of
the work has been done. AVM and FD-A thank K.\ Slevin for fruitful
discussions. Part of the calculations were performed at the Aula Sun Cluster
and the Cl\'{u}ster de C\'{a}lculo de Alta Capacidad para T\'{e}cnicas
F\'{i}sicas, funded by the UCM and the UE under FEDER programme. Work in
Madrid was supported by EC (project MOSAICO) and BSCH-UCM (project
PR34/07-15916).

\end{document}